# A "Janus" double sided mid-IR photodetector based on a MIM architecture

Mario Malerba[1*,], Mathieu Jeannin[1], Stefano Pirotta[1], Lianhe Li[2], Alexander Giles Davies[2], Edmund Linfield[2], Adel Bousseksou[1], Jean-Michel Manceau[1] and Raffaele Colombelli[1,#]

[1] *Centre de Nanosciences et de Nanotechnologies (C2N), CNRS UMR 9001, Universite Paris-Saclay, 91120 Palaiseau, France*
[2] *School of Electronic and Electrical Engineering, University of Leeds, Woodhouse Lane, Leeds LS2 9JT, United Kingdom*

**ABSTRACT**

We present a mid-IR (λ ~ 8.3 µm) quantum well infrared photodetector (QWIP) fabricated on a mid-IR transparent substrate, allowing photodetection with illumination from either the front surface or through the substrate. The device is based on a 400 nm-thick GaAs/AlGaAs semiconductor QWIP heterostructure enclosed in a metal-insulator-metal (MIM) cavity and hosted on a mid-IR transparent ZnSe substrate. Metallic stripes are symmetrically patterned by e-beam lithography on both sides of the active region. The detector spectral coverage spans from λ~7.15 µm to λ~8.7 µm by changing the stripe width $L$ – from $L$=1.0 µm to $L$=1.3 µm – thus frequency-tuning the optical cavity mode. Both micro-FTIR passive optical characterizations and photocurrent measurements of the two-port system are carried out. They reveal a similar spectral response for the two detector ports, with an experimentally measured $T_{BLIP}$ of ~200K.

**Introduction**

Quantum well infrared photodetectors (QWIPs) are amongst the most suitable detectors for imaging, high speed and/or heterodyne detection in the mid-infrared spectral range (mid-IR, λ = 3 – 30 µm). Applications cover trace gas detection, atmospheric studies or space science, IR imaging, free-space optical communications [1-4]. They combine a high sensitivity akin to Mercury Cadmium Telluride (MCT) or InSb detectors, with an ultra-fast intrinsic response time [5]. Recently, the use of resonant metallic cavities has both enhanced their sensitivity and allowed room-temperature operation at the long wavelength end of the mid-IR spectral range [4, 6-7]. Their large electronic bandwidth makes them also excellent candidates for mid-IR cameras.

To date, cavity-embedded QWIPs make use of resonant metal-insulator-metal (MIM, or metal-metal) cavities enclosing the semiconductor active region between a continuous metallic plane and patterned metallic nanostructures, that act simultaneously as resonant subwavelength antennas and as cavity wall. The continuous metallic plane is a technological requirement originating from the fabrication procedure of the device, relying up to now on Au-Au thermo-compression wafer bonding. It presents two main drawbacks: (i) it induces additional parasitic capacitances reducing the electronic bandwidth of the detectors [6], and (ii) it blocks optical access from the backside of the sample. While the first drawback has recently been circumvented using fabrication-intensive dry-etching techniques, the second one is *by design* impossible to avoid. Recently, in view of developing cavity-enhanced ISB devices, original metal-metal architectures that enable bonding on arbitrary, potentially transparent substrates have been introduced [8] and also 3D structuring of the semiconductor active region [9-11].

In this letter we present a mid-IR QWIP based on metallic microcavities symmetrically patterned on both sides of the semiconductor heterostructure and fabricated on a mid-IR transparent ZnSe substrate. It permits detection with illumination from both sides of the device and – potentially – visible illumination from the backside (being ZnSe also transparent for visible wavelengths larger than 500 nm). We optically characterize the device and show that absorption within the optical cavity exhibits the same spectral features, as expected, of a system on a continuous metallic ground plane. We measure the photocurrent response and current-voltage (I-V) characteristics for different cavity resonance frequencies and different temperatures. Our architecture is thus directly compatible with the integration of RF coplanar lines without having to etch through a thick metallic layer [4]. It has a clear potential for heterodyne applications where collinear alignment between the local oscillator and the input signal beams can be difficult or unwanted [12]. The efficient backside injection of the local oscillator beam allows an independent optical path for the signal beam impinging on the device from the front side. Finally, it enables the integration with high-speed, chip level, multiplexed electronic readout circuits towards compact, high resolution, ultrasensitive and ultrafast mid-IR cameras.

**Device Fabrication**

The QWIP active region is composed of 8 periods of epitaxially-grown square quantum wells (**Al$_{0.25}$Ga$_{0.75}$As**/GaAs **25**/5.2 nm), Si-doped to a nominal $n_{2D} = 3 \times 10^{11}$ cm$^{-2}$ and designed to operate around $\lambda = 8.4$ μm (heterostructure design adapted from [6]). The active region is embedded between contact layers (50 nm and 100 nm of GaAs doped to a nominal $n_{3D} = 4 \times 10^{18}$ cm$^{-3}$, as top and bottom contacts, respectively), and is separated from the GaAs wafer by a 400-nm-thick Al$_{0.60}$Ga$_{0.40}$As etch-stop layer.

Figure 1(a) describes the device fabrication. It consists of a first patterning/metal evaporation step, followed by the transfer of the AR from its original substrate to a new mid-IR transparent host, a new aligned patterning/metal deposition step, and a final dry-etching which defines the structures. In detail, a first electron beam lithography is performed on the native sample to define stripes of *width L* spanning between 0.9 μm and 1.35 μm, with a fixed period P = 4 μm, on areas of approximately 80x80 μm² (Fig. 1b and 1c). Next, Ti (5 nm) and Au (150 nm) are evaporated and lifted off, forming the first metal/semiconductor contact. A second step of optical lithography is carried out, to connect all the devices to a common "ground" bonding pad. A Ti/Au/Ti (5/200/20 nm) stack is evaporated and lifted off. The final 20nm thick Ti layer is used to enhance the bonding strength of the metallic pads in the following step. The sample is then flipped and bonded to a host 1-mm-thick, optical-grade polished ZnSe substrate using a commercial epoxy (Epotek 353ND). The native GaAs substrate is then selectively etched in a citric acid solution until the $Al_{0.6}Ga_{0.4}As$ etch stop layer. The latter is removed in HF, exposing the pristine doped layer. A second EBL lithography is performed with a careful alignment of the same design on the previous pattern, followed by Ti/Au (5/150 nm) metal evaporation. We estimate the lateral alignment error to be around 20 nm. A final photolithography step and metallization are carried out to define individual bonding pads, allowing each device to be electrically addressed and measured as a single detecting "pixel". The non-metallized, exposed semiconductor regions are finally etched in a RIE-ICP reactor using a $SiCl_4$/Ar plasma, using the gold pattern/contacts as mask. Finally, an $O_2$ plasma is used to remove the thin epoxy layer remaining between the grating fingers.

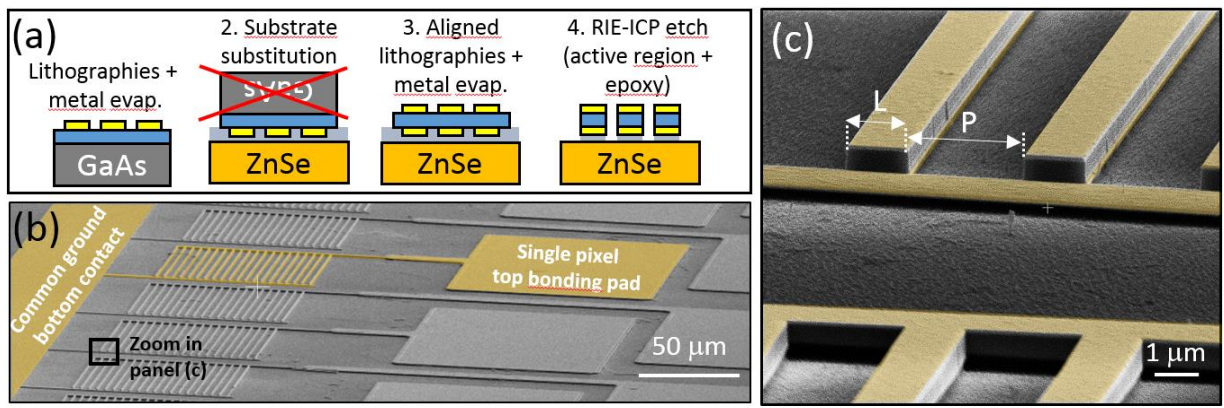

Figure 1 - (a) Main steps for sample fabrication : the active region is sandwiched in a symmetric MIM layout, and bonded to a mid-IR transparent ZnSe substrate. (b) and (c) colorized SEM images showing the sample layout: a common electrode provides ground connection to all the devices, while a dedicated bonding pad selectively addresses each pixel. The stripe width (*L*) varies between 0.9 μm and 1.35 μm, while the distance *P* is kept constant

Note: for technological simplicity, we did not employ alloyed ohmic contacts. The Ti/Au contacts on the n+ GaAs layers thus introduce a Schottky barrier that possibly degrades the overall QWIP performance with respect to state-of-the-art devices. This can be overcome using low-temperature (200°C) annealed Pd/Ge/Ti/Au ohmic contacts on GaAs instead [11, 13-14].

Overall, the architecture thus consists of a symmetric MIM microcavity, embedding an ISB active region, fabricated on a mid-IR transparent support as shown in the two scanning electron microscopy (SEM) images presented in Figs. 1(b) and 1(c). The *modal* properties of the system are identical to a conventional MIM system on a continuous ground plane. The novelty is in the electromagnetic coupling properties: the transparent substrate allows addressing the devices from either the top surface (hereafter called *the front side*), or through the substrate (*the back side*). Borrowing the term Janus from the ancient roman mythological god *Ianus*, depicted as having two faces looking in opposite directions – its name directly connected to the meaning of "doorway" – and the port terminology from the TCMT (temporal coupled mode theory) formalism, we define the two illumination sides as two excitation and/or collection *ports* for the light.

**Micro-FTIR characterization**

The devices were first optically characterized using a mid-IR microscope coupled to a Fourier Transform Infrared (FTIR) Spectrometer (Nicolet Nexus 870). The microscope permits to perform micro-reflectance (R) and micro-transmission (T) by precisely selecting the area probed by the mid-IR beam using a pair of illumination and collection slits along the light path. Reflectivity and transmissivity spectra were obtained at room temperature, focusing and collecting light only on the 100x100 μm² area defined by the grating with two aligned 32x Cassegrain objectives (see sketches in fig. 2). Light impinges on the sample with an angle of incidence of 25-35° and is TM-polarized using a KRS5 polarizer along the optical path. The grating stripes are orthogonal to the plane of incidence. A control spectrum is collected in the same conditions on nearby, un-patterned ZnSe areas (dotted gray spectra in panels a-f of figure 2), to experimentally evaluate the reflectivity at the air/ZnSe interface.

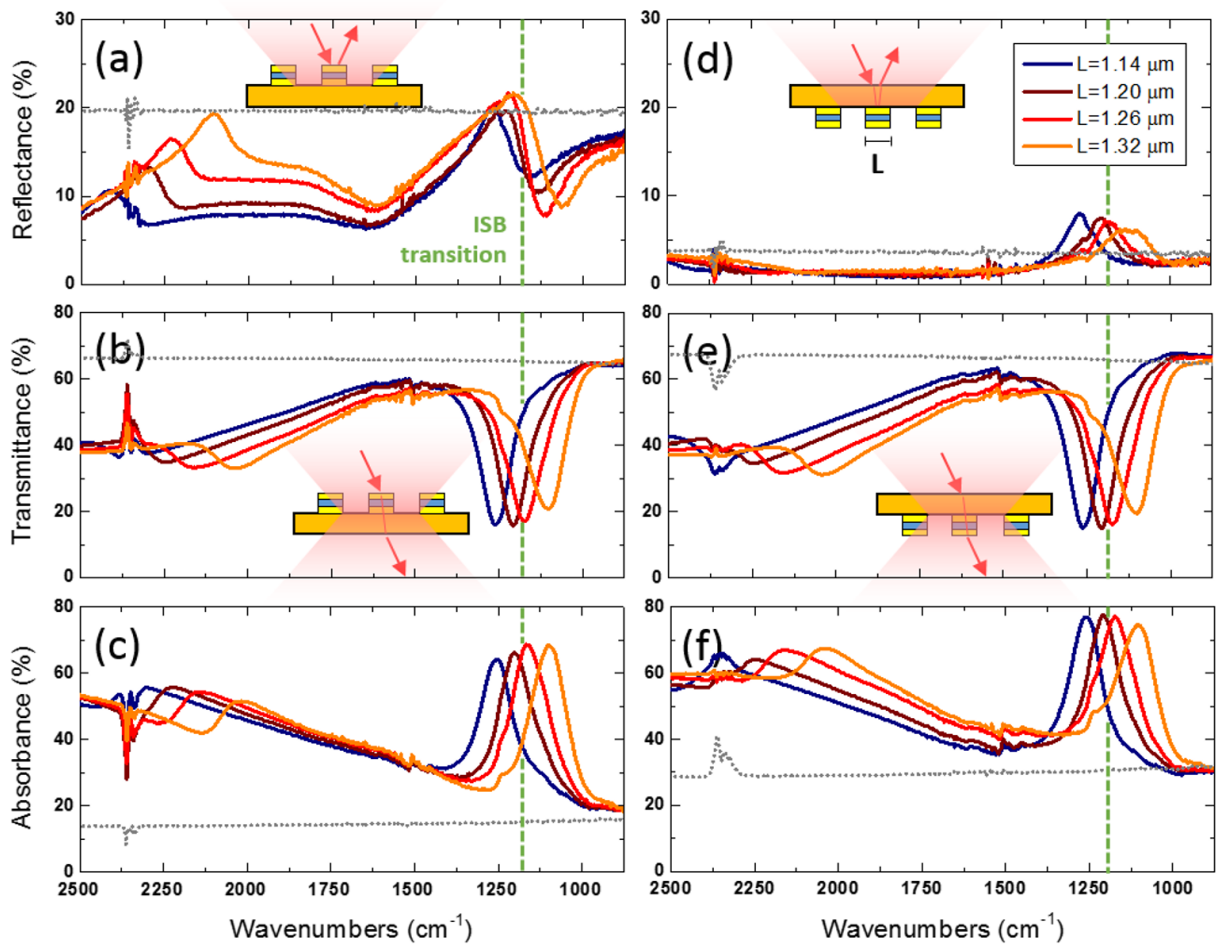

Figure 2 - Micro-FTIR passive optical characterizations for different optical cavity geometries (stripe width L varies to span the photonic resonance across the ISB transition at 1180 cm$^{-1}$). TM-polarised light is focused directly on the MIM structures with a Cassegrain objective from the top-side, collecting (a) reflectivity and (b) transmission. The same measurements are performed by rotating the sample upside down and impinging through the ZnSe substrate, focalizing at the interface hosting the active region ((d) and (e)). The absorption spectra are inferred as A=1-R-T (panels (c) and (f)). Dotted gray lines: spectra collected on the bare ZnSe substrate, in nearby regions; green vertical line: ISB transition (1176 cm$^{-1}$)

Figures 2(a) and 2(b) show respectively the room-temperature reflectance and transmittance spectra of the MIM cavities for a selection of different stripe widths (L = 1.14 μm, 1.20 μm, 1.26 μm and 1.32 μm) with light incident from the front side. Pronounced resonant features are observed in both reflection and transmission, correctly frequency-tuning with the cavity at the change of L. We focus on the low-energy resonances that correspond to the $TM_{01}$ mode of the microcavities [15-17] that are relevant for the operation, between 1000 cm$^{-1}$ and 1400 cm$^{-1}$. The reflectivity spectra exhibit a characteristic dispersive lineshape, while transmission spectra exhibit a pronounced dip, with a contrast of ~40%. These lineshapes are characteristic of two-port systems [18].

From R and T, we can extract the front-side illumination absorbance as A = 1-R-T (Fig. 2(c)). The resonances of interest show an absorption magnitude of around 45%. Note that these spectra also display a large background absorption of around 20% in the region of interest. The origin of this background, that is not expected to occur in a transparent substrate, can be explained by analyzing the optical losses in spectra taken on non-patterned ZnSe (gray dots in Fig. 2). We measure a spectrally independent radiation loss of the same order of magnitude: as there is in principle very little absorption in the ZnSe, we assign the effect to multiple internal reflections or scattering within the transparent substrate slab. Due to the finite aperture of the Cassegrain objective and the truncation of the light path by the microscope slits, part of the reflected/transmitted light is not delivered to the detector. Overall, at the cavity resonance, we measure an absorbance of around 45% once corrected for the background. Considering that the maximum absorption value for a thinner-than-λ, two-port system illuminated from only one side is 50%, the developed detector proves very efficient in energy harvesting [19-20].

The same set of measurements was performed with back-side illumination: reflectance, transmittance and absorbance are shown in Figures 2(d-f), with very similar results. As expected by the Helmholtz reciprocity principle, the transmittance spectra are practically identical, while the reflectance differs from the front-side configuration [21-22]. An absorbance as high as 45% is measured in this back-side configuration too, after correction for the light lost by geometrical effects.

The passive optical characterizations confirm that photons can be efficiently injected in our device from both sides, with very similar performances as far as the absorbed power is concerned.

**DC electrical characterization**

We now turn to the DC opto-electronic characterization of the devices for both front-side and back-side illuminations. The sample is mounted on a copper block with a central hole allowing optical access also from the backside and it is fixed on the cold finger of a liquid nitrogen-cooled, continuous-flow cryostat. It is surrounded by a cryo-shield with a 24° circular field of view (FOV) access that can be opened to expose the device to a room temperature (300 K) black body radiation or closed to characterize its electrical response in the dark. Current-voltage characteristics as a function of the sample's temperature were measured using a Keithley 2461 source meter.

We present in Fig. 3 the dark current (black solid lines), and the background current (measured by exposing the device to a 300 K black body either through the front side – red dashed line – or the back side – blue dashed line). The device shown here has a stripe width L = 1.23 µm that sustains a cavity mode at $\lambda_{res}$ = 1180 cm$^{-1}$, in resonance with the ISB transition ($\lambda_{ISB}$ = 1176 cm$^{-1}$). Additional data for an off-resonant device (L = 1.11 µm) is presented in the Supplementary Information (SI-1). As noted previously, the metallic contacts form a Schottky barrier of approx. 0.6 V (shaded gray area in Fig. 3(a)) where the photodetection process is hampered. The Current-voltage I(V) characteristics is asymmetric, due to the difference in processing conditions between the top and bottom contact layers. The background current is larger than the dark current for the three reported temperatures. From the measurement at 78 K, it appears that operation under negative bias is slightly favorable as it yields a larger background to dark current ratio.

By repeating the measurements over a broader temperature range (78 K - 300 K), it is possible to obtain an estimate of the background limited infrared photo-detection temperature ($T_{BLIP}$) as the temperature where $I_{bkg} = 2*I_{dark}$. It represents the minimal cooling temperature of the device so that its performances are only limited by the dark current thermal noise.

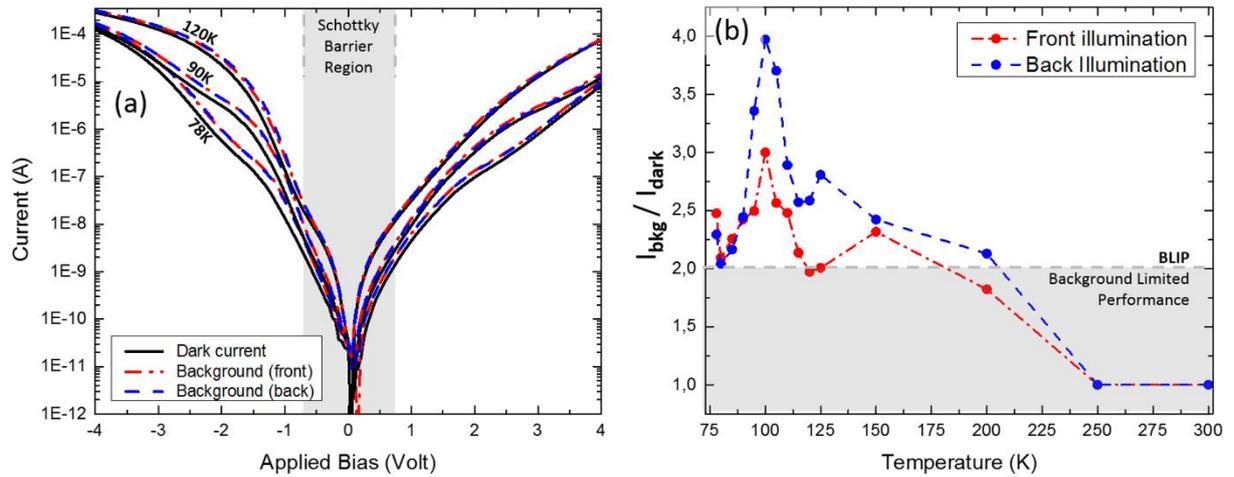

Figure 3 – (a) Current-voltage characteristics of the resonant (L = 1.23 µm) device in the dark (black solid lines), or upon ambient temperature background illumination from the front (blue dashed line) or back (red dashed-dotted line) side. (b) Ratio of the background current to the dark current as a function of temperature, showing that the operation is background limited up to around 200K.

The result is reported in Fig. 3(b), as the ratio of the background current over the dark current for front (red dots and dashed line) and back (blue dots and dashed line) illumination as a function of temperature. For each temperature, the operating bias is selected to maximize the background to dark current ratio, typically in the -1 V to -1.5 V range. *De facto*, this indicates the most favorable operating conditions where the S/N ratio is the largest. From the results in Fig. 3(b) we estimate a $T_{BLIP}$ around 200K for both front and back side configurations. We note that this could be improved in the future with the use of ohmic contacts.

The background to dark current ratio increases slightly with the temperature from 78 K to 115 K before dropping rapidly below 2. This effect has been previously reported in QWIPs operating in the mid-IR and THz [11, 23-24]. It originates from the extra energy necessary to promote electrons from the doped GaAs contact layers above the first QW barrier, which results in a triangular injection band profile. In our case, this phenomenon is further amplified by the presence of the Schottky barriers at the GaAs/Metal interfaces. As a result, electrons need a larger energy to overcome the potential barrier and be emitted/collected from the contacts.

**Photodetection**

We finally discuss the photo-detection response of this two-port "Janus" QWIP. Light coming from the internal source of an FTIR spectrometer (Bruker Vertex 70v) is focused on the device with a parabolic mirror. The sample is kept on the same support previously used, a cryostat protected by a cryo-shield and accessible from the exterior through the same FOV of about 24° to ensure that only one side of the device is exposed to external radiation. A low-noise amplifier (SR570) is used to apply an external bias on the device and amplify its current response with a typical sensitivity of 500 nA/V. The amplified signal is first fed directly into a lock-in amplifier, used to finely align the photodetector pixel on the focused spot of the incoming radiation, and then is sent back to the FTIR as "external detector signal" for spectrum acquisition in rapid scan mode.

Following the scheme of previous experiments, we characterized the detector for the two coupling configurations – frontside and backside illumination – and collected signal from three independent pixels (stripe width L = 1.26 µm, L = 1.14 µm and L = 1.02 µm). To complete the analysis, we collected data at different operating temperatures (for brevity, published in the supporting information section SI-2).

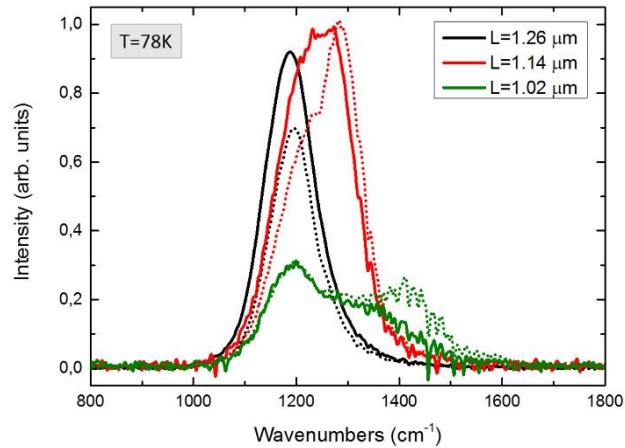

Figure 4 – Photocurrent spectra obtained at 78K for three different devices (the stripe width is indicated in the legend) illuminated from the front (solid lines) or back (dotted lines) surfaces. All spectra are arbitrarily normalized to the best performing condition (L = 1.14 µm with backside illumination).

Figure 4 shows detector performance at 78 K for the three investigated pixels, with photocurrent spectra normalized to the best performing layout. The grating with L = 1.26 µm features a photonic resonance peaked at $\lambda_{res}$ = 1180 cm$^{-1}$, matching the ISB transition at 1176 cm$^{-1}$. It shows a sharp and symmetric photo-detection peak for both front-side and back-side illumination. By decreasing the value of the geometric parameter *L* we explore the off-resonance region, as the detection peak blue-shifts accordingly. An asymmetry in the photocurrent spectrum is introduced by the blue-shift of the photonic mode, that modulates the spectral shape, enhancing the detection at the tail of the ISB transition. A double peaked photo-response is observed for the grating L=1.14 µm, and it is even more apparent for a larger detuning (L = 1.02 µm). Despite the decrease in detectivity, the bandwidth for which photocurrent is generated increases, allowing photo-detection between 1000 cm$^{-1}$ and 1500 cm$^{-1}$. This evidences the large tunability of the photodetection range by tuning the photonic architecture of the device.

**Conclusion**
In conclusion, we have demonstrated the operation of a metal-insulator-metal cavity-enhanced QWIP on a mid-IR transparent substrate, that shows very similar performances in both direct and backside illumination. This technological innovation allows for a vast number of perspective applications that were up to now essentially not available due to the use of Au-Au thermocompression wafer bonding techniques. It is in principle now possible to process the detector in a matrix architecture with flip-chip soldering solutions that permit the use of fast, standardized electronic read-out-circuits to develop a mid-IR, cavity-enhanced QWIP camera. Simultaneous two-side illumination is also a desirable feature in experiments involving e.g. heterodyne detection schemes where a precise alignment and overlap between two beams is necessary. In the case of a single-side illumination sensor, a trade-off must be found between non-colinear alignment or the use of beam-splitters, whereas the use of a two-side illumination device allows simultaneous optical access from both sides and permits to remove the beam-splitter. This approach has been demonstrated in the THz with hot-electron bolometers [12], but never in the mid-IR range of the spectrum. Finally, the performances of our device can be enhanced by using ohmic contact alloys and annealing schemes, as well as more refined cavity architectures like square patch microcavities instead of a 1D stripe geometry.

**Acknowledgements**
We acknowledge financial support from the European Union FET-Open Grant MIRBOSE (737017) and from European Union's Horizon 2020 Research and Innovation Program, under the Marie Skłodowska Curie Grant Agreement No. 748071. We also acknowledge financial support from the French National Research Agency (project "IRENA" and "SOLID"). This work was partly supported by the French RENATECH network.

**Data availability**
Meaningful data are provided in the text and in the supporting information. The data that support the findings of this study are available from the corresponding author upon reasonable request.

# SUPPORTING INFORMATION

Mario Malerba[1*,], Mathieu Jeannin[1], Stefano Pirotta[1], Lianhe Li[2], Alexander Giles Davies[2], Edmund Linfield[2], Adel Bousseksou[1], Jean-Michel Manceau[1] and Raffaele Colombelli[1,#]

[1] Centre de Nanosciences et de Nanotechnologies (C2N), CNRS UMR 9001, Universite Paris-Saclay, 91120 Palaiseau, France
[2] School of Electronic and Electrical Engineering, University of Leeds, Woodhouse Lane, Leeds LS2 9JT, United Kingdom

**SI-1 – DC electrical characterization of an off-resonance layout.**

For the sake of completeness, and to allow a comparison with the best performing resonator array (L=1.23 μm, optically resonant with the ISB transition), we show the data collected for a geometry where the optical cavity mode is not superposed to the ISB transition, (L=1.11 μm, hosting an optical cavity mode at $\lambda_{res}$=1300 cm$^{-1}$).

With a mismatch between $\lambda_{RES}$ and $\lambda_{ISB}$ of approximately 130 cm$^{-1}$, we record that the $T_{BLIP}$ falls below 78K, which implies a non-negligible drop in detectivity (fig. SI-2). This is not surprising, however, as the device is not optimized for minimizing dark current, or in pursuit of high-T performance. As for the on-resonance geometry presented in the main text, also for this configuration we see an increase in detectivity around 100K, connected to the non-annealed character of the contacts, which requires a slight increase in the available energy to overcome the Schottky barrier.

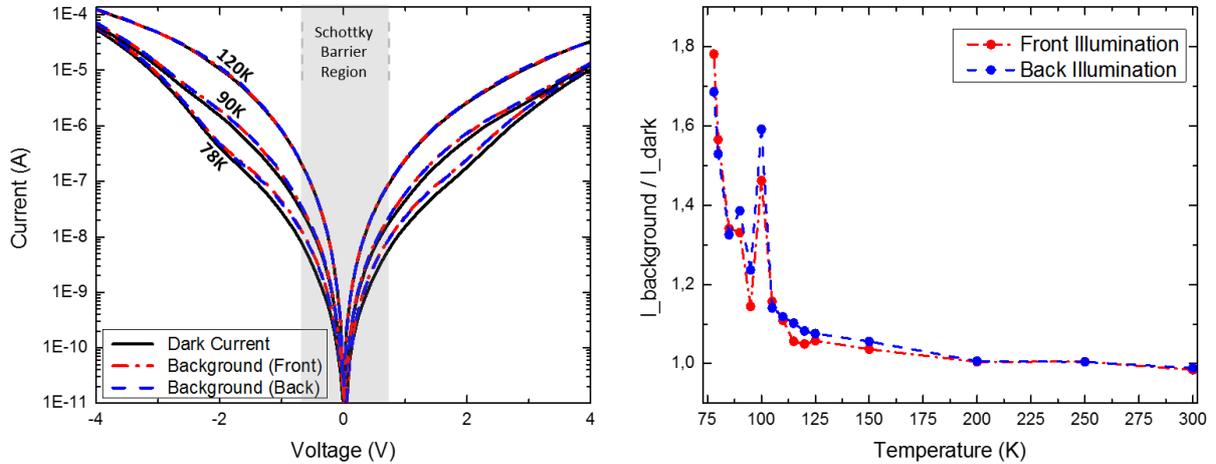

*Figure SI-1* - (left) Current-voltage characteristics of the off-resonance (L = 1.11μm) device in the dark (black solid lines), or upon ambient temperature background illumination from the front (blue dashed line) or back (red dashed-dotted line) side. (b) Ratio of the background current to the dark current as a function of temperature, showing that the operation is background-limited already at liquid nitrogen temperature.

**SI-2 – Photocurrent spectra at different operating temperatures.**

To validate the data presented in figure 3 (main text), we collected spectra at different temperatures from the same *on-resonance* photodetector pixel. As for the I(V) curves and the other photocurrent measurements in the main text, also in this case the sample was illuminated from one single port (frontside), while the remaining surrounding regions were protected by a cryo-shield. We see a trend in photocurrent which retraces the trend in $I_{background} - I_{dark}$ (plot in figure 3b, main text), with an increase in detection from 78K, exhibiting a maximum around 100K and a progressive decrease when operating at higher temperatures. All the measurements were recorded using the FTIR in "rapid scan" mode. A weak signal was detected up to 150K, here not shown for clarity.

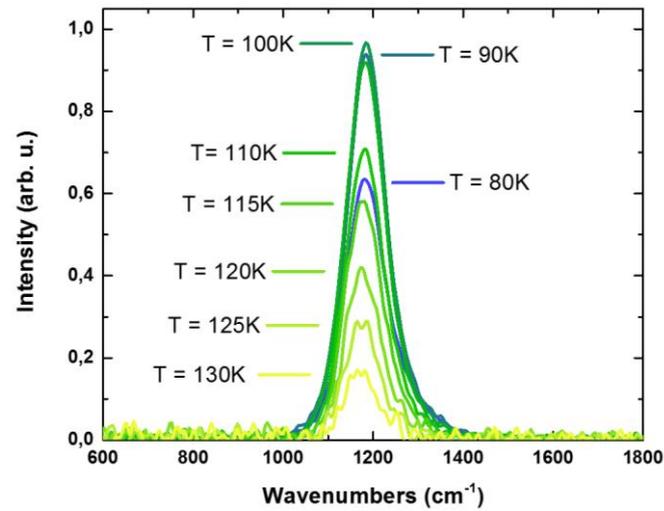

Figure SI-2 – Photocurrent spectra at different operating temperatures for the on-resonance geometry (pixel L=1.23μm), illumination impinging from the frontside. Data is normalized to unity, taking the highest value as reference.